\documentclass[aps,prl,twocolumn,superscriptaddress,showpacs,amsmath,amssymb]{revtex4-1}

\usepackage[dvipdfmx]{graphicx} 

\usepackage{dcolumn}% Align table columns on decimal point
\usepackage{bm}% bold math
\usepackage{color}

\begin{document}

\preprint{APS/123-QED}

\title{Energy of the $^{229}$Th Nuclear Clock Isomer Determined by Absolute $\gamma$-ray Energy Difference}

\author{A. Yamaguchi}
\affiliation{Quantum Metrology Laboratory, RIKEN, Wako, Saitama 351-0198, Japan}
\author{H. Muramatsu}
\altaffiliation[Present address: ]{National Aeronautics and Space Administration/Goddard Space Flight Center}
\affiliation{Institute of Space and Astronautical Science/Japan Aerospace Exploration Agency, Sagamihara, Kanagawa 252-5210, Japan}
\author{T. Hayashi}
\affiliation{Institute of Space and Astronautical Science/Japan Aerospace Exploration Agency, Sagamihara, Kanagawa 252-5210, Japan}
\author{N. Yuasa}
\affiliation{Department of Applied Quantum Physics and Nuclear Engineering, Kyushu University, Fukuoka 819-0395, Japan}
\author{K. Nakamura}
\affiliation{Safety and Nuclear Security Administration Department, Japan Atomic Energy Agency, Chiyoda-ku, Tokyo 100-8577, Japan}
\author{M. Takimoto}
\affiliation{Nuclear Fuel Cycle Engineering Laboratories, Japan Atomic Energy Agency, Naka-gun, Ibaraki 319-1194, Japan}
\author{H. Haba}
\affiliation{Nishina Center for Accelerator-Based Science, RIKEN, Wako, Saitama 351-0198, Japan}
\author{K. Konashi}
\affiliation{Institute for Materials Research, Tohoku University, Higashiibaraki-gun, Ibaraki 311-1313, Japan}
\author{M. Watanabe}
\affiliation{Institute for Materials Research, Tohoku University, Higashiibaraki-gun, Ibaraki 311-1313, Japan}
\author{H. Kikunaga}
\affiliation{Research Center for Electron Photon Science, Tohoku University, Sendai, Miyagi 982-0826, Japan}
\author{K. Maehata}
\affiliation{Department of Applied Quantum Physics and Nuclear Engineering, Kyushu University, Fukuoka 819-0395, Japan}
\author{N. Y. Yamasaki}
\affiliation{Institute of Space and Astronautical Science/Japan Aerospace Exploration Agency, Sagamihara, Kanagawa 252-5210, Japan}
\author{K. Mitsuda}
\affiliation{Institute of Space and Astronautical Science/Japan Aerospace Exploration Agency, Sagamihara, Kanagawa 252-5210, Japan}

\date{\today}
% It is always \today, today,
%  but any date may be explicitly specified

\begin{abstract}
The low-lying isomeric state of $^{229}$Th provides unique opportunities for high-resolution laser spectroscopy of the atomic nucleus. We determine the energy of this isomeric state by taking the absolute energy difference between the excitation energy required to populate the 29.2-keV state from the ground-state and the energy emitted in its decay to the isomeric excited state. A transition-edge sensor microcalorimeter was used to measure the absolute energy of the 29.2-keV $\gamma$-ray. Together with the cross-band transition energy (29.2~keV$\to$ground) and the branching ratio of the 29.2-keV state measured in a recent study, the isomer energy was determined to be 8.30$\pm$0.92~eV. Our result is in agreement with latest measurements based on different experimental techniques, which further confirms that the isomeric state of $^{229}$Th is in the laser-accessible vacuum ultraviolet range. 
\end{abstract}

\maketitle

\begin{figure}
\includegraphics[width=1 \linewidth]{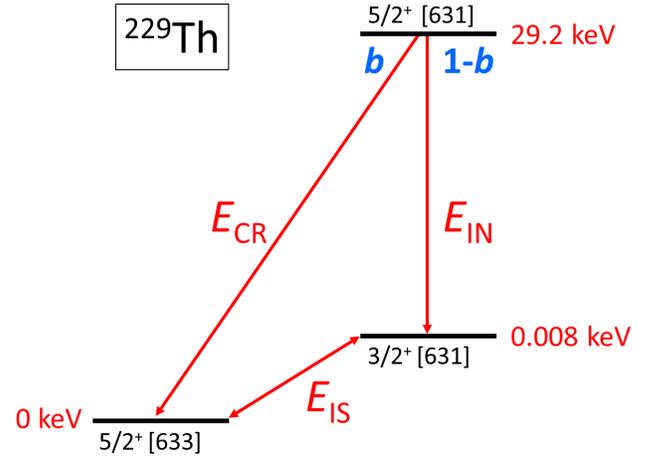}
\caption{\label{Fig1}First three nuclear states of $^{229}$Th are shown with energies. Two rotational bands are labeled by the Nilsson asymptotic quantum numbers. From the 29.2-keV state, the nucleus decays via the cross-band transition ($E_{\rm{CR}}$, 5/2$^+$[631]$\to$5/2$^+$[633]) and inband transition ($E_{\rm{IN}}$, 5/2$^+$[631]$\to$3/2$^+$[631]) with the branching ratios $b$ and $1-b$, respectively.}
\end{figure}

The energy of the first-excited isomeric state of $^{229}$Th is sufficiently low so that it can be excited by laser light \cite{Beck-PRL-2007, Seiferle-Nature-2019}. The natural linewidth of the transition between the ground and the isomer states is predicted to be on the order of mHz \cite{Tkalya-PRC-2015, Minkov-PRL-2017}. Therefore this nuclear transition offers unique opportunities for laser spectroscopy of the atomic nucleus. One of the promising applications is an optical nuclear clock: an atomic clock referencing a nuclear transition \cite{Peik-EL-2003}. Since the atomic nucleus is highly isolated from the environment due to shielding by the electron cloud, fractional accuracy of the nuclear clock is expected to approach 1$\times$10$^{-19}$ \cite{Campbell-PRL-2012}. Extensive experimental efforts have been made to accurately measure this nuclear transition energy ($E_{\rm{IS}}$  in Fig.~\ref{Fig1}) \cite{Beck-PRL-2007, Seiferle-Nature-2019, Wense-Nature-2016, Jeet-PRL-2015, Yamaguchi-NJP-2015, Stellmer-PRC-2018, Masuda-Nature-2019}. While early measurements suggested energy values of 3.5 \cite{Helmer-PRC-1994} or 5.5 eV \cite{Guimaraes-PRC-2005}, in the latest measurements the value drastically increased to 7.8 \cite{Beck-PRL-2007} and 8.28 eV \cite{Seiferle-Nature-2019}. Further measurements based on different experimental techniques are important to improve the confidence in the energy of the isomer state.

Here we determine $E_{\rm{IS}}$ by taking the difference of absolute energies between two 29.2-keV transitions ($E_{\rm{CR}}$ and $E_{\rm{IN}}$ in Fig.~\ref{Fig1}). The cross-band transition energy $E_{\rm{CR}}$ and the branching ratio $b$ were recently experimentally determined by resonantly exciting the 29.2-keV state with narrow-band synchrotron radiation \cite{Masuda-Nature-2019}. In order to determine $E_{\rm{IN}}$, we performed spectroscopy of $\gamma$-rays emitted from the decay of the 29.2-keV state. Two $\gamma$-rays, $E_{\rm{IN}}$ and $E_{\rm{CR}}$, were not directly resolved due to insufficient detector resolution. Instead, a single spectrum, whose peak energy $E_{\rm{IN}}^{\rm{obs}}$ is the weighted average of the two peaks ($E_{\rm{IN}}^{\rm{obs}} = b E_{\rm{CR}} + (1-b) E_{\rm{IN}}$), was observed. Thanks to the precise value of $E_{\rm{CR}}$ and $b$ \cite{Masuda-Nature-2019}, we were able to extract $E_{\rm{IN}}$ and determine $E_{\rm{IS}}$ by $E_{\rm{CR}}-E_{\rm{IN}}$.

The $E_{\rm{IN}}^{\rm{obs}}$ value was previously measured by using a low energy photon spectrometer \cite{Barci-PRC-2003} and X-ray spectrometer \cite{Beck-PRL-2007}. In this study, we measured $E_{\rm{IN}}^{\rm{obs}}$ by using a single-pixel transition edge sensor (TES) microcalorimeter \cite{Irwin_CPD_2005}. We find that our $E_{\rm{IN}}^{\rm{obs}}$ value disagrees with the previous measurement (3.8$\sigma$ discrepancy from the value in Ref. \cite{Barci-PRC-2003} where $\sigma$ is their standard deviation), but the extracted $E_{\rm{IS}}$ is in agreement with latest two measurements based on different experimental techniques \cite{Beck-PRL-2007, Seiferle-Nature-2019}.

\begin{table}[tb]
 \caption{\label{Table:CalibrationPeaks} The $\gamma$- and X-rays used for the energy calibration are shown with its energy and natural linewidth in the unit of eV. Natural linewidth for $^{241}$Am $\gamma$-line is negligible for the present study.}
  \begin{ruledtabular}
  \begin{tabular}{lllc}
    Line  & Energy & Width & Ref. \\
       \hline
    Ag $K_{\alpha2}$        &  21990.30 (10)   & 9.32    & \cite{Deslattes-RMP-2003, Krause-JPCRD-1979} \\
    Ag $K_{\alpha1}$        &  22162.917 (30)  & 9.16    & \cite{Deslattes-RMP-2003, Krause-JPCRD-1979} \\
    $^{241}$Am(26.3~keV)    &  26344.6 (2)   & -      & \cite{Helmer-NIMPRA-2000} \\
    Cs $K_{\alpha2}$        &  30625.40 (45) & 15.80   & \cite{Deslattes-RMP-2003, Krause-JPCRD-1979} \\
    Cs $K_{\alpha1}$        &  30973.13 (46) & 15.60  & \cite{Deslattes-RMP-2003, Krause-JPCRD-1979} \\
  \end{tabular}
 \end{ruledtabular}
\end{table}

We detected the 29.2-keV $\gamma$-rays emitted following $\alpha$-decay of $^{233}$U. The 26~MBq of $^{233}$U was first chemically purified as $^{233}$UO$_2$Cl$_2$ by an ion exchange column to remove daughter nuclei. It was then dissolved in dilute hydrochloric acid and sealed in a container made of 0.5-mm-thick fluorocarbon polymer, resulting in attenuation of the 29.2-keV $\gamma$-rays by the container wall of less than 5\%. Diameter and thickness of this $^{233}$U source is 25~mm and 3~mm, respectively. The source is attached to outside of a 1-mm thick beryllium window of a dilution refrigerator in which the TES is installed. The distance between the source and the TES pixel is 5 cm. The TES pixel is made of a titanium-gold bilayer whose transition temperature is designed to be 164~mK. The 3.6-$\mu$m thick and 300-$\mu$m square gold absorber is attached to the TES pixel \cite{Muramatsu_IEEE_2017}. Based on results of Monte Carlo simulations, we estimate the total detection efficiency of the 29.2-keV $\gamma$-rays, including both solid angle and absorption efficiency of $\gamma$-rays by an absorber, to be on the order of 10$^{-7}$. Energy resolution for 29.2-keV $\gamma$-rays was observed to be 36~eV (full width at half maximum) at a heat sink temperature of 90~mK.

\begin{figure}
\includegraphics[width=1 \linewidth]{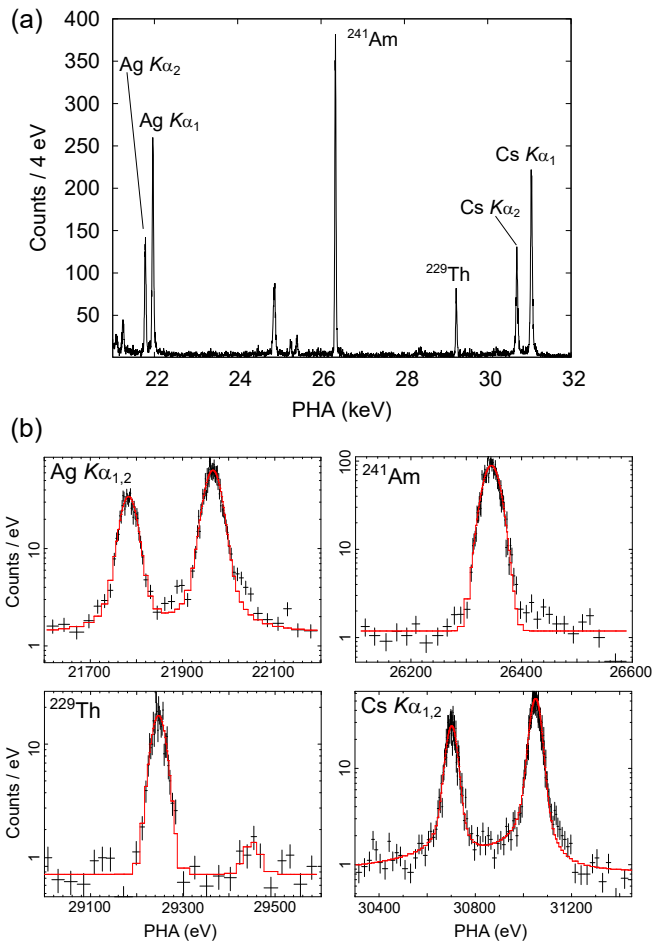}
\caption{\label{Fig2}(a) The $\gamma$-ray and X-ray PHA spectra of $^{233}$U source around 29~keV obtained with a TES. The abscissa is a PHA value which is scaled to approximately the expected energy scale prior to calibrating the non-linearity. Five peaks used for energy calibration and a $\gamma$-ray peak corresponding to decay from the 29.2-keV state in $^{229}$Th are labeled. (b) Detailed spectra of each calibration line and the 29.2-keV line are shown. Red curve represents the maximum likelihood estimation result. Energy resolution for the 29.2-keV line was measured to be 36~eV (full width at half maximum).}
\end{figure}

The TES is biased with a pseudo constant voltage using a shunt resistor \cite{Irwin_CPD_2005}, and the signal is obtained from the TES current measured by a superconducting quantum interference device (SQUID) array amplifier \cite{Sakai_JLTP_2014}. The SQUID output voltage was recorded by a 15-bit digitizer and, in the offline analysis, restored to the TES current. The pulse-height of the TES current exhibits a non-linear response to the incident $\gamma$-ray energy because of several reasons \cite{Peille-SPIE-2016}. A part of the non-linearity can be removed by converting the TES current to the TES resistance value \cite{Bandler_NIMPRA_2006}. Spectral data was collected for 18 consecutive days, during which the detector gain drifted due to a change of the detector temperature. We thus divided all data into blocks, where the gain drift is negligible. Each data block was processed independently with the optimum filter \cite{Szymkowiak-JLTP-1993} to make a pulse height amplitude (PHA) spectrum. The total PHA spectrum (Fig.~\ref{Fig2}) was obtained by combining all PHA spectra.

The PHA shows non-linearity to the $\gamma$-ray energy, and thus energy calibration plays a crucial role in this study. We chose five $\gamma$- and X-rays as the energy calibration lines around the 29.2-keV target peak (as labeled in Fig.~\ref{Fig2}(a)). In Table~\ref{Table:CalibrationPeaks}, the energy and natural linewidth of each calibration line are listed. These calibration peaks are simultaneously recorded with the 29.2-keV target peak by attaching $^{241}$Am (2.4~MBq) and $^{133}$Ba (59~kBq) sources behind the $^{233}$U source. The Cs $K_{\alpha1,2}$ X-rays are emitted in the process of electron capture (EC) decay of $^{133}$Ba. The 26.3-keV $\gamma$-rays are from $\alpha$-decay of $^{241}$Am. Since our $^{241}$Am source is electroplated onto a silver plate, the Ag $K_{\alpha1,2}$ X-rays are emitted mainly when the strong 59.5-keV $\gamma$-rays from the $^{241}$Am source are photoelectrically absorbed by the silver plate.

\begin{table}[tb]
 \caption{\label{Table:Chi-square} Comparisons of goodness of the fits for linear, quadratic and cubic polynomial calibration curves.}
  \begin{ruledtabular}
  \begin{tabular}{cccc}
  $n ^{a}$    &   dof & $\chi^2$/dof & $F$ \\
       \hline
    1       & 3     & 10909   &   \\
    \noalign{\vspace{-4pt}} & & & 30563 \\
    2       & 2     & 1.07    &    \\
    \noalign{\vspace{-4pt}} & & & 0.05 \\
    3       & 1     & 2.05    &   \\
  \end{tabular}
 \end{ruledtabular}
\raggedright  $^{a}$ The order of polynomial.  The number of free parameters  is  $n+1$.
\end{table}

The centroid PHA values of all calibration peaks and the 29.2-keV target peak were determined by model fitting of the PHA spectrum with a maximum likelihood method. For X-ray calibration peaks, we used a Voigt function as a model function. The natural linewidth was fixed to the literature value (see Table~\ref{Table:CalibrationPeaks}) which was converted into the PHA value by assuming a locally linear PHA-to-energy relation \cite{Muramatsu_IEEE_2017}. The $\gamma$-ray peaks at 26.3~keV ($^{241}$Am) and 29.2~keV ($^{229}$Th) were fitted with a Gaussian function. Here we assumed a constant background within a fitting region. Detailed spectra are shown in Fig.~\ref{Fig2}(b) where the red curve represents the best-fit model curve. The 1$\sigma$ errors of the centroid PHA values were also estimated from the model fits. The total counts in the 29.2-keV peak was 630(30). A small peak observed at the higher-PHA side of the 29.2-keV target peak in the $^{229}$Th spectrum corresponds to the 7/2$^+$[631](71.82 keV)$\to$7/2$^+$[633](42.43 keV) cross-band transition (not shown in Fig.~\ref{Fig1}).

\begin{figure}
\begin{center}
\includegraphics[width=1 \linewidth]{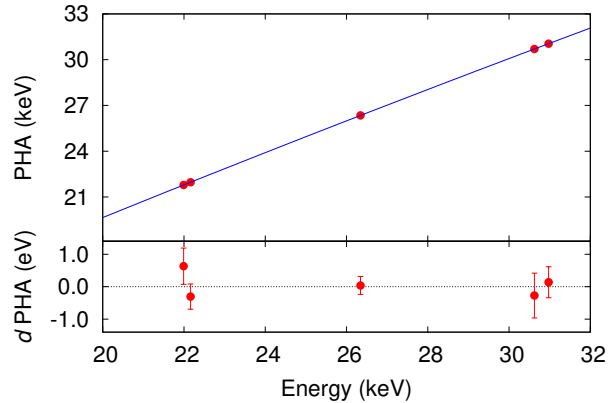}
\end{center}
\caption{\label{Fig3}The PHA values of five calibration lines are plotted as a function of their energies. The result of $\chi^2$ fitting with a 2nd-order polynomial function is shown by a blue curve and corresponding $d$PHA values are shown below. The 1$\sigma$ errors of PHA values are smaller than the symbols of the upper panel.}
\end{figure}

In order to accurately determine the 29.2~keV $\gamma$-ray energy, we need to define a suitable calibration curve which precisely converts PHA to energy. Within the limited energy range from 21.9~keV to 31~keV, in which all calibration lines and the 29.2-keV target line are contained, we found that the calibration curve can be well-approximated by a polynomial function. To find the optimal polynomial order, we performed $\chi^2$ fitting of the PHA-to-energy relation with $n$-th order polynomial functions where $n=$1, 2, or 3. For the $\sigma$s of the $\chi^2$, the 1$\sigma$ errors of the centroid PHA values from the maximum-likelihood PHA fits were used. The errors of energies will be included later in the systematic errors. The best $n$ value was determined to be 2 from the $F$ value of $\chi^2$ improvement \cite{Bevington-1969} (Table~\ref{Table:Chi-square}). The $F$ value from $n$-th to $(n+1)$-th order polynomial model is defined by $F(n\to n+1) = [\chi^2(n) - \chi^2(n+1)]/[\chi^2(n+1)/{\rm dof}(n+1)]$, where $\chi^2(n)$ and dof$(n)$ are, respectively, the minimum $\chi^2$ value and the degrees of freedom of $n$-th order polynomial fit. $F(1\to2)$ is very large and the improvement is statistically significant. On the other hand, $F(2\to3)$ is smaller than unity, which indicates that the added parameter just represents the statistical fluctuation. We thus employed a 2nd-order polynomial function as a calibration curve. In Fig.~\ref{Fig3}, the PHA values of five calibration lines are plotted as a function of their energies. The best-fit 2nd order polynomial calibration curve is shown by a blue curve. The $d$PHA (= data $-$ model) values at each calibration peak are also shown in Fig.~\ref{Fig3}. By using this calibration curve, the absolute energy of the 29.2-keV $\gamma$-ray $E_{\rm{IN}}^{\rm{obs}}$ was determined to be 29182.51~eV. For further validation of the calibration curve, we checked the energy of the 25.5-keV Ag $K_{\beta2}$ X-ray and the 25.3-keV $^{229}$Th $\gamma$-ray lines in Fig.~\ref{Fig2}(a), neither of which were used to define the calibration curve. The first energy, for Ag in a metal form, was determined to be 25456.6(11)~eV. Although there is some uncertainty of less than 0.4 eV mainly due to the so-called shake-off effect \cite{Hansen-AISP-1985}, the value is in agreement with 25456.71(31)~eV in Ref.~\cite{Deslattes-RMP-2003}. The second energy was determined to be 25308.4(19)~eV, which agrees with 25310.6(8)~eV in Ref.~\cite{Helmer-PRC-1994}.

As statistical error for $E_{\rm{IN}}^{\rm{obs}}$, we considered the following two errors. One is the 1$\sigma$ error of centroid PHA value of the 29.2-keV spectrum converted to energy, which is 0.67~eV. The other is caused by 1$\sigma$ statistical errors of three coefficients of the 2nd order polynomial calibration curve. To evaluate this, we calculated $\chi^2$ values for all combinations of three coefficients which satisfied $\chi^2 \le \chi^2_{\rm{min}}+1$, where $\chi^2_{\rm{min}}$ is the minimum $\chi^2$ value of the fit. We converted the centroid PHA of 29.2-keV line to the energy for each combination. Then the minimum and the maximum energies give the error range. The total statistical error was estimated from root sum square of the two errors, which is 0.72~eV.

The following three errors are considered as systematic errors for $E_{\rm{IN}}^{\rm{obs}}$. The first is the errors in the literature values as listed in Table~\ref{Table:CalibrationPeaks}. The second is the errors due to various effects which could shift the X-ray energies. For such effects, we regarded hyperfine, chemical and shake-off effects as follows \cite{Hansen-AISP-1985}. The final state of the $K_{\alpha}$ X-ray transition has two hyperfine states \cite{Breit-PR-1930}. When the transition probability to these states is not equal, the X-ray spectral peak shifts. Such a shift could occur when, for example, X-rays are generated by the EC decay \cite{Borchert-PL-1977}. Since whether X-rays used in the literature were generated by photoionization or EC decay is not clear, we took half of the hyperfine splitting energy as the largest possible error due to hyperfine effect, which is estimated to be 0.03~eV for Ag $K_{\alpha1,2}$ and 0.36~eV for Cs $K_{\alpha1,2}$ \cite{Hansen-AISP-1985}. The chemical effect can be partially estimated based on the formula given in Ref.~\cite{Sumbaev-JETP-1970}. However, information on the chemical condition of both our calibration sources and sources used in the literature is sparse. We therefore did not include the chemical effect in our systematic error. If we assume the largest possible coordination number (12 for Cs and 8 for Ag \cite{Shannon-AC-1976}) and electronegativity difference (2.65 for Cs and 2.05 for Ag) as an example, the chemical shift could be estimated to be 0.08 and 0.10~eV for Ag $K_{\alpha1,2}$, 0.26 and 0.31~eV for Cs $K_{\alpha1,2}$, which would contribute to the total systematic error by only 0.01 eV. For the error induced by the shake-off effect \cite{Crasemann-PRC-1979}, we conservatively estimated it to be 0.05~eV for all X-rays based on the discussions in Ref.~\cite{Hansen-AISP-1985}. The combined systematic error caused by the aforementioned first and second errors was determined by a Monte Carlo simulation where we fixed the PHA value for all calibration lines and scanned the energy within each systematic error. Here we used a Gaussian function as a distribution of errors for the literature value and a flat function for all other errors. At each set of energies, we calculated a calibration curve and derived the 29.2-keV energy. The standard deviation of the 29.2-keV energies was taken as the combined systematic error, which is 0.22~eV. The third error is the error due to an inappropriate functional form for the calibration curve. A cubic spline curve with natural boundary conditions \cite{Fowler-Metrologia-2017} was tested as a calibration curve, resulting in an energy discrepancy of 0.25~eV. We therefore determined the overall systematic error by taking a root sum square of both errors (0.22~eV and 0.25~eV), which is 0.33~eV.

\begin{figure}
\includegraphics[width=1 \linewidth]{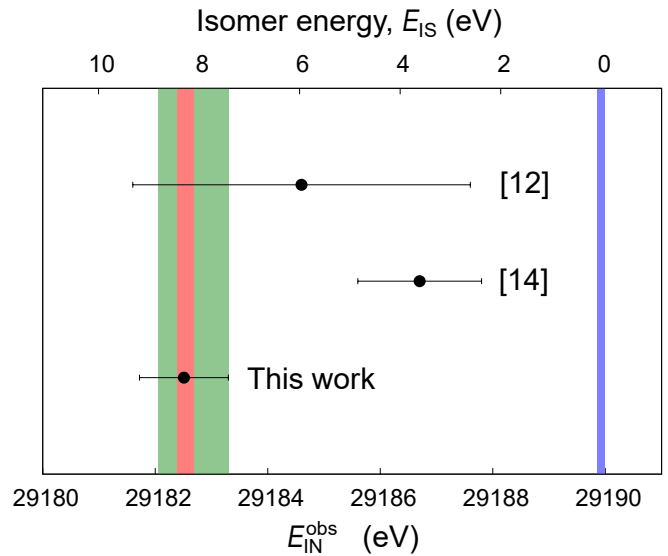}
\caption{\label{Fig4} Comparison of $E_{\rm{IN}}^{\rm{obs}}$ in this study with those in Refs.\cite{Helmer-PRC-1994, Barci-PRC-2003}. Error bars for our $E_{\rm{IN}}^{\rm{obs}}$ represent root sum square of statistical and systematic error. The blue bar represents $E_{\rm{CR}}$ with its 1$\sigma$ error. The isomer energy calculated by using $E_{\rm{CR}}$ and $b$ \cite{Masuda-Nature-2019} is shown in the upper side of the figure. The green and red bands represent the isomer energy reported in Refs.~\cite{Beck-PRL-2007} and \cite{Seiferle-Nature-2019}, respectively, with their corresponding uncertainties, where the isomer energy for Ref.~\cite{Beck-PRL-2007} is re-estimated by replacing $b$ with the experimental value reported in Ref.~\cite{Masuda-Nature-2019}.}
\end{figure}

As an absolute energy $E_{\rm{IN}}^{\rm{obs}}$, we concluded
\begin{eqnarray*}
E_{\rm{IN}}^{\rm{obs}}=29182.51~\pm~0.33~(\rm{sys})~\pm~0.72~(\rm{stat})~eV.  
\end{eqnarray*}
The root sum square of statistical and systematic errors is 0.79~eV. In order to determine $E_{\rm{IS}}$ by $E_{\rm{CR}}-E_{\rm{IN}}$, we need to extract $E_{\rm{IN}}$ from $E_{\rm{IN}}^{\rm{obs}}$. As discussed earlier, they are related to each other by $E_{\rm{IN}}^{\rm{obs}} = b E_{\rm{CR}} + (1-b) E_{\rm{IN}}$. The $E_{\rm{CR}}$ and $b$ values were experimentally measured to be 29189.93$\pm$0.07~eV and 0.106$\pm$0.027 in a recent study \cite{Masuda-Nature-2019}. Combining these values, we determined the isomer energy to be
\begin{eqnarray*}
E_{\rm{IS}}=8.30~\pm~0.45~(\rm{sys})~\pm~0.81~(\rm{stat})~eV.  
\end{eqnarray*}
Here we included the errors of  $E_{\rm{CR}}$ and $b$ in the systematic error. The root sum square of statistical and systematic errors is 0.92~eV.

In Fig.~\ref{Fig4}, the $E_{\rm{IN}}^{\rm{obs}}$ value determined in this study is compared to previous measurements. The isomer energy calculated by $E_{\rm{IS}} = (E_{\rm{CR}}-E_{\rm{IN}}^{\rm{obs}})/(1-b)$ is presented in the upper side of the figure. The blue bar indicates $E_{\rm{CR}}$ \cite{Masuda-Nature-2019}. The green and red bands denote the isomer energies with errors reported in Refs.~\cite{Beck-PRL-2007} and  \cite{Seiferle-Nature-2019}, respectively. In Ref.~\cite{Beck-PRL-2007}, the isomer energy is determined to be 7.8$\pm$0.5 eV by using an estimated $b$. If we replace the $b$ by the experimental value reported in Ref.~\cite{Masuda-Nature-2019}, their isomer energy can be estimated to be 8.1$\pm$0.7 eV, which is shown in Fig.~\ref{Fig4} as a green band. While our new $E_{\rm{IN}}^{\rm{obs}}$ differs from the most accurate previous value (29186.7$\pm$1.1 eV \cite{Barci-PRC-2003}) by more than 3.8$\sigma$ in their uncertainty, the extracted isomer energy agrees with both latest measurements (red and green bands) within their 1$\sigma$ uncertainty. 

In summary, the energy of the first isomeric state in $^{229}$Th is determined to be 8.30$\pm$0.92~eV by measuring the absolute energy difference between two transitions from the 29.2-keV second-excited state. The absolute energy of the 29.2-keV $\gamma$-ray following $\alpha$-decay of $^{233}$U was measured by a single-pixel TES microcalorimeter. Agreement between our $E_{\rm{IS}}$ value and two latest values measured by different experimental techniques further confirms that the energy of the $^{229}$Th nuclear clock isomer is in the laser-accessible vacuum ultraviolet range and paves the way for high-resolution laser spectroscopy of the atomic nucleus.

We thank D. Aoki for supports to conduct this research and T. Yamamura for technical supports to prepare the $^{233}$U source. We also thank T. Masuda, A. Yoshimi, K. Yoshimura and A. Hinton for their valuable comments on the manuscript. This work was supported by JSPS Grant-in-Aid for Scientific Research (B) Grant No. JP18H01241.  A.Y. acknowledges Technology Pioneering Projects in RIKEN. The $^{233}$U sample used in this study is provided by the $^{233}$U cooperation project between JAEA and the Inter-University Cooperative Research Program of the Institute for Materials Research, Tohoku University (proposal no.17K0204). H.M. is partially supported by JSPS Grant-in-Aid for JSPS Fellows Grant No. 17J07990. We fabricated the TES microcalorimeter partly using the nano-electronics fabrication facility of JAXA. The SQUID array amplifier was fabricated by CRAVITY of AIST.


\begin{thebibliography}{99}
\bibitem{Beck-PRL-2007}
   B. R. Beck, J. A. Becker, P. Beiersdorfer, G. V. Brown, K. J. Moody, J. B. Wilhelmy, F. S. Porter, C. A. Kilbourne, and R. L. Kelley, Energy Splitting of the Ground-State Doublet in the Nucleus $^{229}$Th, Phys. Rev. Lett. \textbf{98}, 142501 (2007); B. R. Beck, C. Y. Wu, P. Beiersdorfer, G. V. Brown, J. A. Becker, K. J. Moody, J. B. Wilhelmy, F. S. Porter, C. A. Kilbourne, R. L. Kelley, Improved Value for the Energy Splitting of the Ground-State Doublet in the Nucleus $^{229\rm{m}}$Th, LLNL-PROC-415170 (2009).
\bibitem{Seiferle-Nature-2019}
    B. Seiferle, \textit{et al.}, Energy of the $^{229}$Th nuclear clock transition, Nature \textbf{573}, 243 (2019).
\bibitem{Tkalya-PRC-2015}
    E. V. Tkalya, C. Schneider, J. Jeet, and E. R. Hudson, Radiative lifetime and energy of the low-energy isomeric level in $^{229}$Th, Phys. Rev. C \textbf{92}, 054324 (2015).
\bibitem{Minkov-PRL-2017}
    N. Minkov and A. P\'alffy, Reduced Transition Probabilities for the Gamma Decay of the 7.8 eV Isomer in $^{229}$Th, Phys. Rev. Lett. \textbf{118}, 212501 (2017).
\bibitem{Peik-EL-2003}
    E. Peik and Chr. Tamm, Nuclear laser spectroscopy of the 3.5 eV transition in Th-229, Europhys. Lett. \textbf{61}, 181 (2003).
\bibitem{Campbell-PRL-2012}
    C. J. Campbell, A. G. Radnaev, A. Kuzmich, V. A. Dzuba, V. V. Flambaum, and A. Derevianko, Single-Ion Nuclear Clock for Metrology at the 19th Decimal Place, Phys. Rev. Lett. \textbf{108}, 120802 (2012).
\bibitem{Wense-Nature-2016}
    L. von der Wense \textit{et al.}, Direct detection of the $^{229}$Th nuclear clock transition, Nature \textbf{533}, 47 (2016).
\bibitem{Jeet-PRL-2015}
    J. Jeet, C. Schneider, S. T. Sullivan, W. G. Rellergert, S. Mirzadeh, A. Cassanho, H. P. Jenssen, E. V. Tkalya, and E. R. Hudson, Results of a Direct Search Using Synchrotron Radiation for the Low-Energy $^{229}$Th Nuclear Isomeric Transition, Phys. Rev. Lett. \textbf{114}, 253001 (2015).
\bibitem{Yamaguchi-NJP-2015}
    A. Yamaguchi, M. Kolbe, H. Kaser, T. Reichel, A. Gottwald, and E. Peik, Experimental search for the low-energy nuclear transition in $^{229}$Th with undulator radiation, New J. Phys. \textbf{17}, 053053 (2015).
\bibitem{Stellmer-PRC-2018}
    S. Stellmer, Y. Shigekawa, V. Rosecker, G. A. Kazakov, Y. Kasamatsu, Y. Yasuda, A. Shinohara, and T. Schumm, Toward an energy measurement of the internal conversion electron in the deexcitation of the $^{229}$Th isomer, Phys. Rev. C \textbf{98}, 014317 (2018).
\bibitem{Masuda-Nature-2019}
    T. Masuda \textit{et al.}, X-ray pumping of the $^{229}$Th nuclear clock isomer, Nature \textbf{573}, 238 (2019).
\bibitem{Helmer-PRC-1994}
    R. G. Helmer and C. W. Reich, An excited state of $^{229}$Th at 3.5 eV, Phys. Rev. C \textbf{49}, 1845 (1994).
\bibitem{Guimaraes-PRC-2005}
    Z. O. Guimar{\~a}es-Filho and O. Helene, Energy of the 3/2$^{+}$ state of $^{229}$Th reexamined, Phys. Rev. C \textbf{71}, 044303 (2005).
\bibitem{Barci-PRC-2003}
    V. Barci, G. Ardisson, G. Barci-Funel, B. Weiss, O. El Samad, and R. K. Sheline, Nuclear structure of $^{229}$Th from $\gamma$-ray spectroscopy study of $^{233}$U $\alpha$-particle decay, Phys. Rev. C \textbf{68}, 034329 (2003).
\bibitem{Irwin_CPD_2005}
    K. D. Irwin and G. C. Hilton, Transition-Edge Sensors, in \textit{Cryogenic Particle Detection}, Top. Appl. Phys., vol. 99, edited by C. Enss (Springer, Berlin, Heidelberg, 2005) p. 63. 
\bibitem{Deslattes-RMP-2003}
    R. D. Deslattes, E. G. Kessler Jr., P. Indelicato, L. de Billy, E. Lindroth, and J. Anton, X-ray transition energies: new approach to a comprehensive evaluation, Rev. Mod. Phys. \textbf{75}, 35 (2003).
\bibitem{Krause-JPCRD-1979}
    M. O. Krause and J. H. Oliver, Natural widths of atomic $K$ and $L$ levels, $K\alpha$ X-ray lines and several $KLL$ Auger lines, J. Phys. Chem. Ref. Data \textbf{8}, 329 (1979).
\bibitem{Helmer-NIMPRA-2000}
    R. G. Helmer and C. van der Leun, Recommended standards for $\gamma$-ray energy calibration (1999), Nucl. Instr. and Meth. A \textbf{450}, 35 (2000).
\bibitem{Muramatsu_IEEE_2017}
    H. Muramatsu, T. Hayashi, K. Maehisa, Y. Nakashima, K. Mitsuda, N. Y. Yamasaki, T. Hara, K. Maehata, A Study of X-Ray Response of the TES X-Ray Microcalorimeter for STEM, IEEE Trans. Appl. Supercond. \textbf{27}, 2101204 (2017).
\bibitem{Sakai_JLTP_2014}  
    K. Sakai, Y. Takei, R. Yamamoto, N. Y. Yamasaki, K. Mitsuda, M. Hidaka, S. Nagasawa, S. Kohjiro, and T. Miyazaki, Baseband Feedback Frequency-Division Multiplexing with Low-Power dc-SQUIDs and Digital Electronics for TES X-Ray Microcalorimeters, J. Low Temp. Phys. \textbf{176}, 400 (2014).
\bibitem{Peille-SPIE-2016}
    P. Peille \textit{et al.}, Performance assessment of different pulse reconstruction algorithms for the ATHENA X-ray Integral Field Unit, in \textit{Proceedings of SPIE 9905, Space Telescopes and Instrumentation 2016: Ultraviolet to Gamma Ray, Edinburgh, United Kingdom, 2016,} edited by J-W. A. den Herder, T. Takahashi, and M. Bautz, 99055W.
\bibitem{Bandler_NIMPRA_2006}
    S. R. Bandler, E. Figueroa-Feliciano, N. Iyomoto, R. L. Kelley, C. A. Kilbourne, K. D. Murphy, F. S. Porter, T. Saab, and J. Sadleir, Non-linear effects in transition edge sensors for X-ray detection, Nucl. Instrum. Methods Phys. Res. A \textbf{559}, 817 (2006).
\bibitem{Szymkowiak-JLTP-1993}
    A. E. Szymkowiak, R. L. Kelley, S. H. Moseley, C. K. Stahle, Signal processing for microcalorimeters, J. Low Temp. Phys. \textbf{93}, 281 (1993).
\bibitem{Bevington-1969}
P. R. Bevington and D. K. Robinson, \textit{Data Reduction and Error Analysis for the Physical Sciences}, (McGraw-Hill, Inc., NY, 1969).
\bibitem{Hansen-AISP-1985}
    P. G. Hansen, B. Jonson, G. L. Borchert, and O. W. B. Schult, Mechanisms for energy shifts of atomic $K$ X rays, Atomic Inner-Shell Physics, (Plenum Press, New York, 1985), p. 237.
\bibitem{Breit-PR-1930}
    G. Breit, Possible Effects of Nuclear Spin on X-Ray Terms, Phys. Rev. \textbf{35}, 1447 (1930).
\bibitem{Borchert-PL-1977}
    G. L. Borchert, P. G. Hansen, B. Jonson, H. L. Ravn, O. W. B. Schult, P. Tidemand-Petersson, and ISOLDE Collaboration, Shifts in xenon K X-ray energies following electron-capture beta decay and the role of nuclear hyperfine structure, Phys. Lett. A \textbf{63}, 15 (1977).
\bibitem{Sumbaev-JETP-1970}
    O. I. Sumbaev, The effect of the chemical shift of the x-ray $K_{\alpha}$ lines in heavy atoms. Systematization of the experimental data and comparison with theory, Sov. Phys.-JETP \textbf{30}, 927, (1970).
\bibitem{Shannon-AC-1976}
    R. D. Shannon, Revised Effective Ionic Radii and Systematic Studies of Interatomic Distances in Halides and Chalcogenides, Acta Cryst. \textbf{A32}, 751 (1976).
\bibitem{Crasemann-PRC-1979}
    B. Crasemann, M. H. Chen, J. P. Briand, P. Chevallier, A. Chetioui, and M. Tavernier, Atomic electron excitation probabilities during orbital electron capture by the nucleus, Phys. Rev. C \textbf{19}, 1042 (1979).
\bibitem{Fowler-Metrologia-2017}
    J. W. Fowler \textit{et al.}, A reassessment of absolute energies of the x-ray L lines of lanthanide metals, Metrologia \textbf{54}, 494, (2017). 
\end{thebibliography}
\end{document}